\documentclass[times,twocolumn]{aastex63} 

\newcommand{\Msun}{\mbox{\,$M_\odot$}}

\newcommand{\ltsimeq}{\raisebox{-0.6ex}{$\,\stackrel
	{\raisebox{-.2ex}{$\textstyle <$}}{\sim}\,$}}
\newcommand{\gtsimeq}{\raisebox{-0.6ex}{$\,\stackrel
	{\raisebox{-.2ex}{$\textstyle >$}}{\sim}\,$}}

\newcommand{\EBminusV}{\mbox{$E_{B-V}$}}

\newcommand{\fion}[2]{[{#1}\,{\sc {#2}}]}

\usepackage{color,soul}

\usepackage{graphics,graphicx}
\usepackage{epsfig}

\usepackage{datetime2}

\shorttitle{Copious Dust in V445 Pup}
\shortauthors{Banerjee et al.}
\begin{document}
\title{V445 Puppis -- Dustier than a Thousand Novae}


\author[0000-0002-9670-4824]{D. P. K. Banerjee}
\affiliation{Physical Research Laboratory, Navrangpura, Ahmedabad, Gujarat 380009, India}

\author[0000-0002-3142-8953]{A. Evans}
\affiliation{Astrophysics Group, Keele University, Keele, Staffordshire, ST5 5BG, UK}

\author[0000-0001-6567-627X]{C. E. Woodward}
\affiliation{Minnesota Institute for Astrophysics, University of Minnesota,
116 Church Street SE, Minneapolis, MN 55455, USA}

\author[0000-0002-1359-6312]{S. Starrfield}
\affiliation{School of Earth \& Space Exploration, Arizona State University, 
Box 871404, Tempe, AZ 85287-1404, USA}

\author[0000-0002-3532-5580]{K. Y. L. Su}
\affiliation{Steward Observatory, University of Arizona, 933 North Cherry Avenue, 
Tucson, AZ 85721, USA}

\author[0009-0009-1831-3914]{N. M. Ashok}
\affiliation{Physical Research Laboratory, Navrangpura, Ahmedabad, Gujarat 380009, India}

\author[0000-0003-1892-2751]{R. M. Wagner}
\affiliation{Department of Astronomy, Ohio State University, 140 W. 18th Avenue, Columbus, OH 43210, USA} 
\affiliation{Large Binocular Telescope Observatory, 933 North Cherry Avenue, 
Tucson, AZ 85721, USA}

\correspondingauthor{C.E. Woodward}
\email{chickw024@gmail.com}
\received{2023 May 10}
\revised{2023 May 30}
\accepted{2023 June 16}
\published{To appear in the ApJL}

\begin{abstract}
V445~Puppis, the only known Galactic helium nova, is a unique testbed to verify 
supernova (SN) theories in the single degenerate channel that involve a white dwarf (WD) 
accreting matter from a helium-rich donor. An estimate of the mass 
of the helium shell on the WD is crucial to deciding whether or not it will undergo a SN detonation. 
In this context, this study estimates the dust and ejecta masses in the 2000 November 
eruption of V445~Pup. Subsequent to its outburst,  the star became cocooned in a dust 
envelope.  An analysis of the spectral energy distribution (SED) 
of the dust using infrared data shows that  V445~Pup produced at least $10^{-3}$\Msun\
of dust which is unprecedented for a classical or recurrent nova. The SED can be explained 
by a combination of a cold dust component at $105 \pm 10$~K, mass $(1.9 \pm 0.8) \times 10^{-3}$\Msun, 
and a warm dust component at $255 \pm 10$~K, mass $(2.2 \pm 1.2) \times 10^{-5}$\Msun. 
For a conservative choice of the gas-to-dust mass ratio in the range 10--100, the  mass of 
the ejecta is  0.01--0.1\Msun. Such a high mass range raises the question: why did V445~Pup 
not detonate as a Type~1a SN as is predicted in certain double-detonation 
sub-Chandrasekhar supernovae formalisms? We re-examine  the nature of  V445~Pup and 
discuss its role as a potential SN progenitor. 

\end{abstract}


\section{Introduction}
\label{sec-Intro}

V445 Pup erupted in 2000 reaching a peak $V$ brightness of 8.46~mag on 2000 Nov~29, 
and then slowly declined with a $t_{2} \gtsimeq 100$~days ($t_{2}$ being the elapsed time
to decline 2 mags from peak brightness). Although the  outburst was first 
reported on 2000 December 30 by Kanatsu \citep{2000IAUC.7552....1K} archival 
All Sky Automated Survey \citep[ASAS,][]{1997AcA....47..467P}
records demonstrated the outburst had begun earlier \citep{2010PZ.....30....4G}.  V445~Pup 
appeared to be a slow nova except that the spectra, both in the optical and near-infrared (NIR), 
recorded in the immediate and  post-eruption stages, were unique in not showing the hydrogen 
lines conventionally seen in a nova outburst.  Instead, there were many lines of carbon, helium, 
and other  metals; the C and He lines were specially prominent in the NIR
\citep{2008A&A...482..865I, 2005AIPC..797..647W, 2004AJ....128.2962L, 2003A&A...409.1007A, 2001AJ....122.3313L, 2001IAUC.7717....2W,  2001IAUC.7571....1W}. 
Based on its spectrum \citet{2003A&A...409.1007A} proposed 
V445~Pup to be a helium nova that had undergone a thermonuclear runaway in helium-rich 
matter accreted onto a white dwarf's (WD) surface from a helium-rich 
donor  \citep[e.g.,][]{2003ApJ...598L.107K, 1994ApJ...431..264I}. 

On 2001 Jan 2, $\sim34$ days after peak brightness,  $JHK$ photometry showed that hot 
dust had begun to form \citep{2003A&A...409.1007A}  and 3--14~\micron{} spectroscopy 
obtained on 2001 Jan 31, confirmed the presence of significant 
amounts of carbon dust \citep{2001AJ....122.3313L}. The dust shell rapidly thickened 
from 2001 June and by 2001 October V445~Pup had faded below $V=20$ 
mag \citep{2010PZ.....30....4G}. 

A remarkable $\simeq 2^{\prime\prime}$ hourglass nebula, expanding with time,
(detected around the object with adaptive optics $K_{s}$ band imagery) showed
high velocity outflows \citep{2009ApJ...706..738W}. The knots 
at the tips of the hourglass had velocities as large as  $\sim 8500~\rm{km~s}^{-1}$
\citep{2009ApJ...706..738W}. Flaring radio synchrotron radiation was  persistently 
observed from the object from the beginning and up to 7 years after the outburst 
\citep{2021MNRAS.501.1394N, 2001IAUC.7728....3R}.  It is now clear that this 
non-thermal synchrotron emission was produced from shocks caused by the 
interaction of ejected matter (or a wind) from the WD with a pre-existing equatorial 
density enhancement collimating the ejecta to create the  hour glass nebula 
with its pinched waist \citep{2021MNRAS.501.1394N}.   

In this study, we re-analyze the SED 
of the dust from more recent archival data and conclude that V445~Pup has 
produced an unprecedented amount of dust for a nova. The implications of the large dust 
mass on the role of V445~Pup as a SN Type 1a  progenitor are discussed.   

\section{Distance and Reddening}
\label{sec-DR}

\begin{figure}[ht!]
\figurenum{1}
\begin{center}
\includegraphics[trim=2.5cm 8.2cm 2.5cm 8.4cm, clip, width=0.45\textwidth]{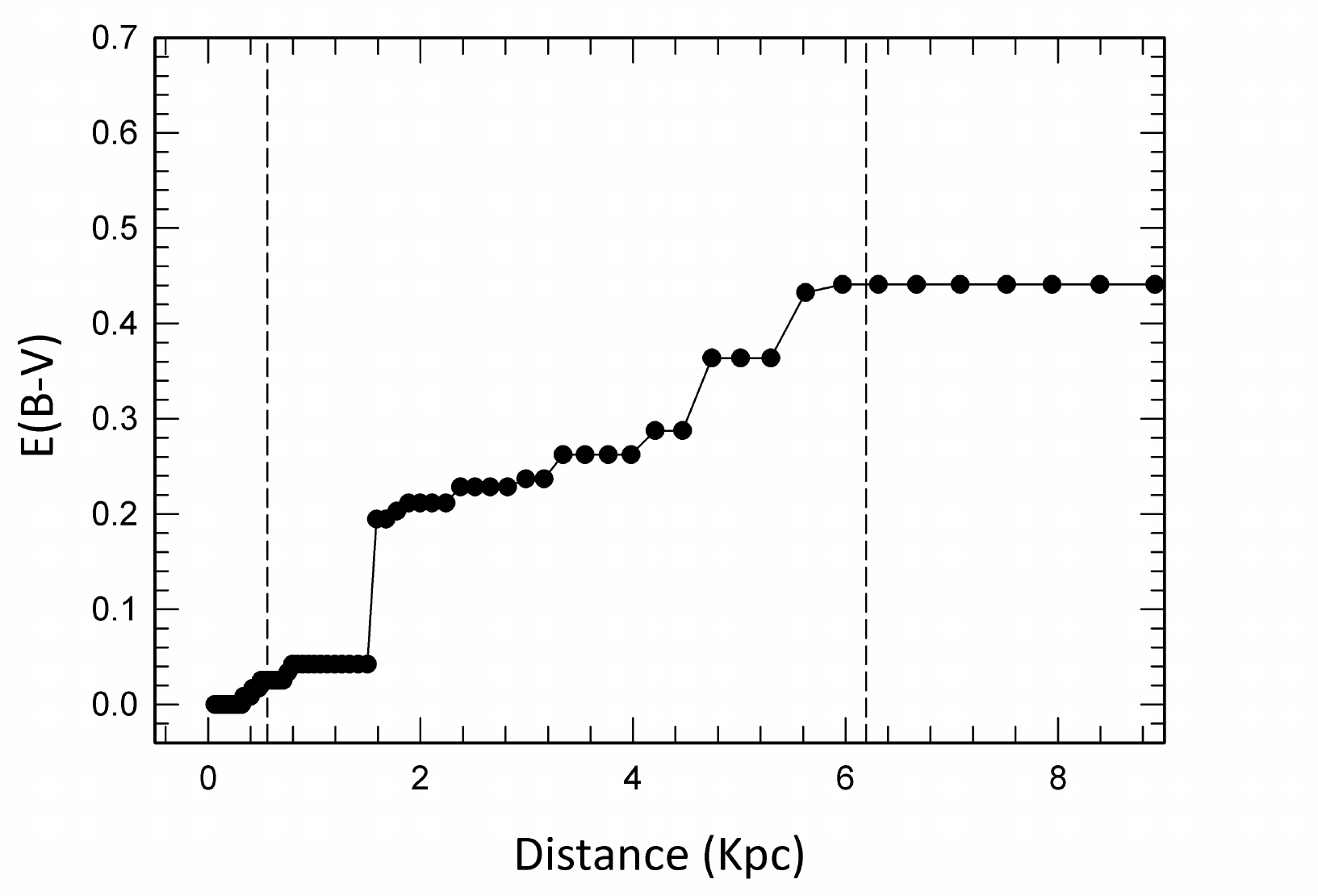}
\caption{Reddening versus distance plot  for V445~Pup from \citet{2019ApJ...887...93G}. The  
$E(g-r)$ values listed therein were converted to a mean \EBminusV using the two relations given 
by the  authors at  \url{http://http://argonaut.skymaps.info/}. The vertical dotted lines 
give the range over which the method is stated to be reliable. See 
text for details.} 
\label{fig:red-dist}
\end{center}
\end{figure}

From the observed expansion parallax of the nebula, \citet{2009ApJ...706..738W} derived a 
distance of $8.2 \pm 0.5$~kpc.  \citet{2008A&A...482..865I} using the radial velocities of the 
Na D1, D2 lines from high dispersion spectra, in conjunction with HI 21cm radio data, 
estimated the reddening and distance to be \EBminusV = 0.51 mag  and 
$3.5 \ltsimeq d(\rm{kpc}) \ltsimeq 6.5.$  We point out that the equivalent 
width of 0.95~\AA{} for the Na DI line in the \citet{2008A&A...482..865I}  data, calibrated 
using the \citet{1994AJ....107.1022R} relations, yields \EBminusV  = 0.45, in reasonable agreement 
with the above.  The \EBminusV = 0.51 derived from the measurements of 
\citet{2008A&A...482..865I} is  consistent with the estimate of \citet{2001IAUC.7571....1W},
\EBminusV $\ltsimeq 0.8$~mag. However, their distance estimate is based on low spatial 
resolution extinction maps of \citet{1980A&AS...42..251N} whereas better data are now available. 
We use reddening data from \citet{2019ApJ...887...93G} to estimate 
the distance (Fig.~\ref{fig:red-dist}). The figure demonstrates, even up to the maximum 
distance of 6.2~kpc (beyond which the data are stated to be unreliable) the \EBminusV 
value has still not reached 0.51 (nor has it reached \EBminusV = 0.45 obtained from the Na DI line).  
This suggests that the distance to V445 Pup is $\gtsimeq 6.2$~kpc.  
We adopt values of  $d = 6.2$~kpc and \EBminusV = 0.51 (hence $A_{V} = 1.6$  
using $A_{V} = 3.1~\times$~\EBminusV).

\section{Analysis and Results}
\label{sec-als}

\subsection{V445 Pup Dust mass}
The SED of the dust was analyzed adopting the formalism of \citet{2016ApJ...817..145S}.
The archival data used for the analysis are presented in Table~\ref{tab:phot-tab}. We assume a 
pure carbon composition for the dust given that the object showed a rich carbon spectrum
in the optical and NIR \citep{2008A&A...482..865I}. Further the 3--14~\micron{} spectrum obtained and modeled by
\citet{2001AJ....122.3313L} showed a featureless continuum, without any silicate or unidentified 
infrared features \citep[UIRs, see][]{2016MNRAS.457.2871E},
and thus strongly favors a carbon composition. 

Assuming an optically thin shell of spherical carbon dust grains with a uniform radius 
$a$, a total mass $M_{i}$ with an equilibrium temperature $T_{i}(\rm{K})$, located at a 
distance $d$ from the observer, the observed flux density $f$ (W~m$^{-2}$~\micron$^{-1}$) is:

\begin{equation}
f_{\nu}^{i} = M_{i} \left( \frac{4 \pi \rho_{(AC)} a^{3}}{3} \right)^{-1}\pi B_{\nu}(\lambda, T_{i}) \, Q_{abs}^{(AC)} \left( \frac{a}{d} \right)^{2}
\label{eq:calflx}
\end{equation}

\noindent where $\rho_{(AC)}$ is the density of amorphous carbon (AC) dust
(1.87~g~cm$^{-3}$) and $Q_{abs}^{(AC)}$ is the absorption efficiency 
of amorphous carbon of radius $a(\micron)$ \citep[BE sample;][]{1996MNRAS.282.1321Z}. 
At IR wavelengths, $Q_{abs}$ $\propto$ $(8\pi a / \lambda)$ and so Eqn.~(\ref{eq:calflx})
becomes independent of the dust grain size $a$ \citep[e.g.,][]{2003pid..book.....K, 1983asls.book.....B}.

The foreground extinction of the emission by silicate grains in the interstellar medium (ISM) 
is taken into account by multiplying the equation above with the exponential term:

\begin{equation}
exp\biggl\{- \tau_{9.7} \left( \frac{Q_{abs}^{ASil}(\lambda)}{Q_{abs}^{ASil}(9.7~\micron)} \right) \biggr\}.
\label{eq:tau97}
\vspace{0.5mm}
\end{equation}

\noindent We have normalized with the optical depth at 9.7~\micron{} ($\tau_{9.7}$) 
determined from $(A_{V} / \tau_{9.7}) = 18.5 \pm 1.5$ \citep{1984MNRAS.208..481R}
with $A_{V} = 1.6$.

\begin{deluxetable*}{@{\extracolsep{0pt}}lcccr}
\tablenum{1}
\setlength{\tabcolsep}{3pt} 
%
%
\tablecaption{V445 Puppis Archival IR/mm Photometery\tablenotemark{$\star$} \label{tab:phot-tab}}
\tablehead{
\colhead{Facility} &\colhead{$\lambda$} &\colhead{Brightness} &\colhead{Flux} &\colhead{Epoch}\\ 
& \colhead{($\mu$m)}  & (Jy/mag) &\colhead{(W m$^{2}$ $\mu$m$^{-1}$)}  
 }
\startdata
WISE & 3.4 & $10.39 \pm 0.23$ (mag)    & $5.7080 \times 10^{-15}$ & 2010\\
WISE & 4.6 & $7.297 \pm 0.020$ (mag)  & $2.9105 \times 10^{-14}$ & 2010\\
WISE\tablenotemark{$\dagger$} & 12.0 & $0.554 \pm 0.027$(mag) & $3.9112 \times 10^{-13}$ & 2010\\
WISE\tablenotemark{$\dagger$}  & 22.0 & $-1.316 \pm 0.010$ (mag) & $1.7106 \times 10^{-13}$ & 2010\\
AKARI-IRC & 18.0 & $35.39 \pm 1.82$ (Jy) & $3.2746 \times 10^{-13}$ & 2006\\
AKARI-FIS2 & 65.0 & $12.55 \pm 0.445$ (Jy) & $8.9051 \times 10^{-15}$ & 2006\\
AKARI-FIS2 & 90.0 & $8.023\pm 0.236$ (Jy) & $2.9694 \times 10^{-15}$ & 2006\\
AKARI-FIS2 & 140.0 & $3.430\pm 0.623$ (Jy) & $5.2460 \times 10^{-16}$ & 2006\\
AKARI-FIS2 & 160.0 & 2.710 (Jy) & $3.1736 \times 10^{-16}$ & 2006\\
Spitzer-IRAC\tablenotemark{$\ddagger$}& 3.6 & $0.11\pm  0.003$ (Jy) & $2.540 \times 10^{-14}$ & 2005\\
Spitzer-IRAC\tablenotemark{$\ddagger$} & 4.5 & $ 0.32\pm 0.019$  (Jy) & $4.750 \times 10^{-14}$ & 2005\\
Spitzer-MIPS\tablenotemark{$\ddagger$} & 70.0 & $7.53\pm 0.038)$ (Jy) & $4.603 \times 10^{-15}$ & 2005\\
Herschel & 70.0 & $7.582\pm 0.064$ (Jy) & $4.6388 \times 10^{-15}$ & 2012\\
Herschel & 160.0 & $1.260\pm 0.044$ (Jy) & $1.3071 \times 10^{-16}$ & 2012\\
SEST & 1200.0 & $0.0295\pm 0.0054$ (Jy) & $6.1400 \times 10^{-20}$ & 2003\\
\enddata
\tablecomments{\, $^{\star}$Data retrieved from various mission archives hosted at the
NASA/IPAC Infrared Science Archive including Spitzer (doi: 10.26131/IRSA433), 
WISE (doi: 10.26131/IRSA142),  Herschel (doi: 10.26131/IRSA79), and
AKARI (doi:10.26131/IRSA180, 10.26131/IRSA181).
$^{\dagger}$The fractional pixel saturation in the 12 and 22~\micron{} fluxes are 
0.28 and 0.12 respectively but saturation effects are corrected by using profile fitted magnitudes 
(\url{https://wise2.ipac.caltech.edu/docs/release/allsky/expsup/sec6_3d.html}).\, $^{\ddagger}$ 
Spitzer data reductions described in \citet{2020ApJ...898...21S}.
The AKARI fluxes are averages over multiple detections made during the mission lifetime 
between 2006-2007.
}
\end{deluxetable*}

\begin{figure}[!ht]
\figurenum{2}
\begin{center}
\includegraphics[trim=1.5cm 8.2cm 2.5cm 7.6cm, clip, width=0.5\textwidth]{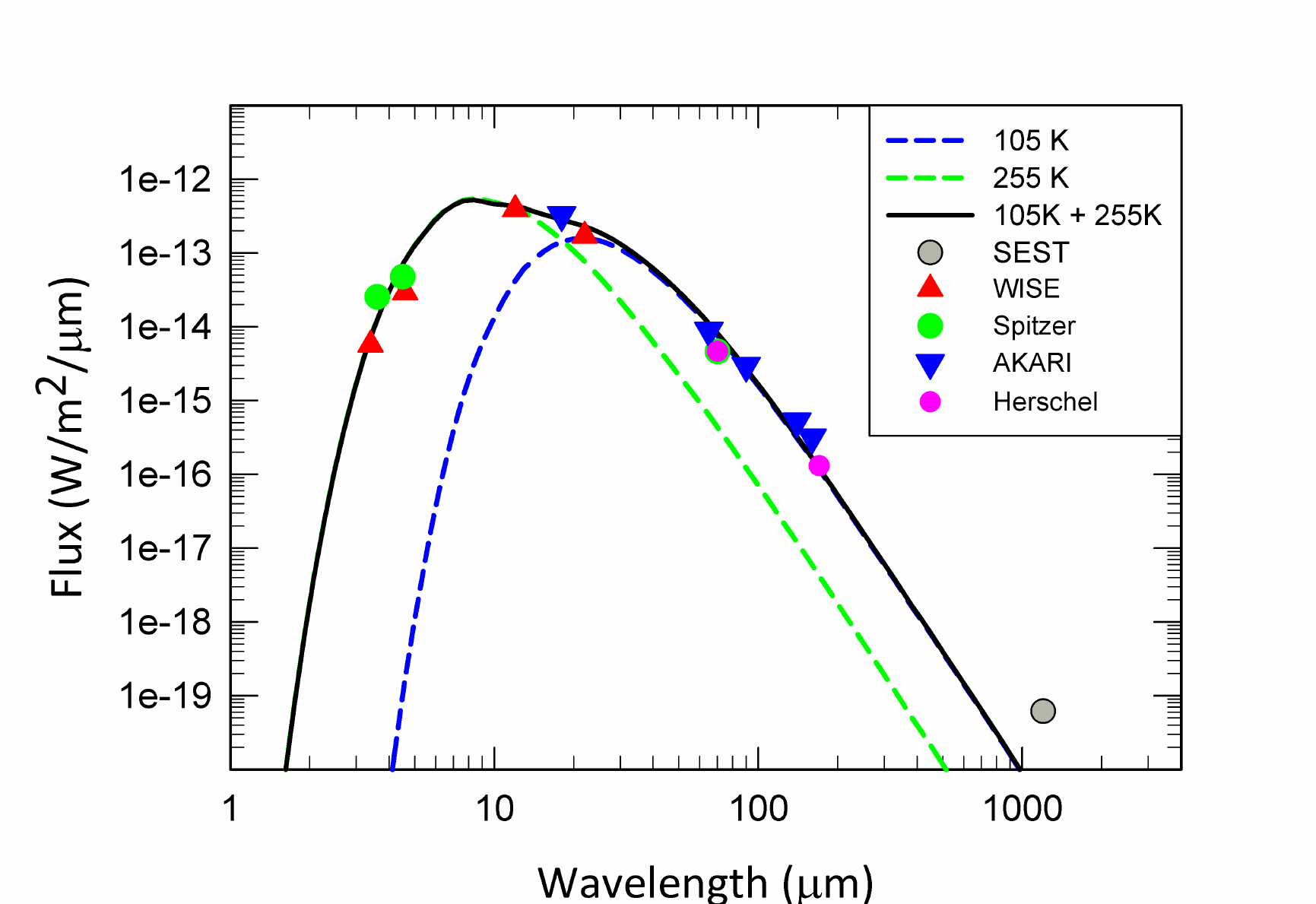}
\caption{Model fit to the SED of V445~Pup using amorphous carbon 
grains \citep[BE sample;][]{1996MNRAS.282.1321Z}. The black bold line is the composite 
of a 105~K  plus a 255~K component. The 1.2mm point was not considered 
in the fitting.} 
\label{fig:fig-sed-1} 
\end{center}
\end{figure}

The SED can be  understood as a combination of two components of amorphous carbon dust. 
The first component at $105 \pm 10$~K has a mass $(1.9 \pm 0.8) \times 10^{-3}$\Msun. The second is a 
warm component, $255 \pm 10$~K, with a mass $(2.2 \pm 1.2) \times 10^{-5}$\Msun. 
The decomposition of the SED is shown in Fig.~\ref{fig:fig-sed-1}. To cross-check the mass 
estimates, the  $Q_{abs}$ of ACAR sample \citep{1996MNRAS.282.1321Z} were also used, yielding 
very similar results as the BE sample. We have not considered the 
1.2~mm point in the fits as it is unclear whether the mm continuum flux is due 
to dust or  from free-free emission from ionized gas (as discussed later). 

Clearly there is a large mass of cool dust, based largely on the SED modeling of the long 
wavelength ($\lambda \gtsimeq 10$~\micron) photometry. Were the dust shell 
optically thick (e.g., $\tau \gtsimeq 5$) at these wavelengths, rather than optically
thin as assumed, then that would require an $A_{V} \simeq 50$ for 
extinction $\propto \lambda^{-1},$ comparable to that seen in the most opaque molecular clouds
and likely is not reasonable. It is outside the scope of this study to compute the dust mass for more 
complex geometries (e.g., a bipolar morphology with ad-hoc assumptions on the amount of  
equatorial material enhancement). 

Our SED fits involve modeling of data that is not contemporaneous. However, 
the main result of the paper - the large dust mass - is largely based on fitting the 
longer wavelength ($\gtsimeq 10$~\micron) AKARI, Herschel, WISE and Spitzer 
data taken between 2005-2012. These data are well-fit by our model suggesting 
that the dust temperature did not change significantly between the different epochs, 
increasing our confidence that the data at these wavelengths are not significantly variable.
There is some variability in the WISE and Spitzer 3.6~\micron{} and 4.5~\micron{} data, but the hotter
dust component which is used to fit this data contributes only a small percent of the dust mass.
So the total dust mass estimate should be reasonably reliable.

A similar but brief study, estimating the dust mass in V445~Pup was conducted 
by \citet{2017PKAS...32..109S}. However, their modeling was limited to only the AKARI data 
with no data short-ward of 9~\micron{} to constrain the Wien side of the SED. The source of the 
9~\micron{} AKARI flux used by \citet{2017PKAS...32..109S} is also unclear. It is not listed in the master records in the AKARI 
database\footnote{\url{https://darts.isas.jaxa.jp/astro/akari/data/AKARI-IRC_Catalogue_AllSky_PointSource_1.0.html}.} 
The mm flux was not discussed and \citet{2017PKAS...32..109S} invoked an unrealistic 
value of foreground extinction of $A_{V} = 12.5$ to fit the SED. The present modeling is 
hence more comprehensive, improved, and realistic.  \citet{2017PKAS...32..109S} do conclude
that large dust masses are extant, comprised of a combination of cold amorphous carbon (125~K)
with a  mass of  $(0.45^{+0.66}_{-0.27}) \times 10^{-3}$ \Msun\ and warm amorphous 
carbon (250~K) with a mass of ($1.8^{+1.0}_{-0.5}) \times10^{-5}$ \Msun. 

 What becomes evident is that no nova, either recurrent nova (RN) or classical nova (CN), has 
produced as much dust ($\simeq 10^{-3}$\Msun) as V445~Pup (adopting
distances of 3.5~kpc or 8.2~kpc as extrema do not radically change the mass estimate
as the dust mass scales as $d^{2}$). The typical mass of the dust produced 
in a dust producing nova is $10^{-6}$ to $10^{-9}$\Msun{} 
\citep[][]{2022arXiv221112410E}. If a canonical value of gas-to-dust mass 
of 100 is assumed,  the mass of the ejecta could be as large as  0.1~\Msun. This is 
unprecedented. 

If the Swedish-ESO Submillimeter Telescope \citep[SEST,][]{1989A&A...216..315B} 
1.2~mm data point is included in the dust SED fitting (however, see \S\ref{sec:mm_submm}), then 
additional modeling suggests that, apart from the two components used in the present 
analysis, an additional cooler  component at  $\simeq 30$ to 50~K and with a mass 
of $10^{-2}$\Msun{} is required to fit the composite SED. It thus appears certain that the 
mass ejected by V445~Pup was very large and that the accreted shell on the WD
shell at the time of outburst was massive, at least $10^{-2}$\Msun, for  a most conservative choice 
of 10 for the gas-to-dust mass ratio. V445~Pup could have potentially erupted as  a SN~1a, 
which curiously it did not. However, we first discuss the possible origin of the mm flux. 

\subsection{Millimeter/sub-mm studies of novae}
\label{sec:mm_submm}
Millimeter/sub-mm studies of novae appear to be few. The two possible origins for mm 
continuum fluxes are free-free emission \citep[e.g., V1974~Cygni,][]{1992IAUC.5516....3I}
and/or the Rayleigh-Jeans tail of dust emission 
\citep[e.g., V4743~Sgr,][]{2005ASPC..330..483S, 2003A&A...400L...5N}. Reasonably, the assertion 
that the detection of 1.2~mm emission from the nebula around V445~Pup is from free-free emission
has a basis in two arguments. First, the 1.2 mm detection in  V445~Pup was made in 2003 May about  $\sim 885$~d after 
the outburst, at which stage the ionized nebula around V445~Pup was expected to be $\ltsimeq 1^{\prime\prime}$
in diameter \citep[based on image sizes in][]{2009ApJ...706..738W}. Second, the optically thin free-free 
flux $F_{ff}(\lambda)$ (W~cm$^{-2}$ \micron$^{-1}$) at wavelength $\lambda$(\micron) from an ionized gas with 
electron density $n_{e}$(cm$^{-3}$), assumed equal to $n_{i}$, the ion density), at  temperature 
$T(K)$ and occupying a volume $V$(cm$^{3}$) is \citep{2001A&A...380L..13B}:

\begin{equation}
F_{ff}(\lambda) = \frac{2.05 \times 10^{-30} \lambda^{2} z^{2} g T^{-0.5} n_{e} n_{i} V}{4 \pi d^{2}} 
\label{eqn:ffeq}
\end{equation}

\noindent where $g$ is the Gaunt factor (between 0.3 and 0.5), $z$ is the charge 
(2 for an ionized Helium gas) and $d$ (in cm) is the distance. 

As an illustrative example, the observed mm flux 
$F \rm{(1.2~mm)} = 6.14 \times 10^{-24}$ W~cm$^{-2}$ \micron$^{-1}$ can be reproduced 
by considering a singly ionized He nebula of 0.6 arcsecond extent at  $T = 10,000$~K and  $d = 6.2$~kpc, 
with a  $n_{e} = 2 \times 10^{5}$~cm$^{-3}$  \citep[since \fion{O}{iii} 5007\AA{} was seen at around 
this time,][] {2005AIPC..797..647W}. We approximate the density to be less than the
critical density of that line which is $\simeq 6 \times 10^{5}$ cm$^{-3}$. These assumptions 
are reasonable and serve to show that the mm flux is almost certainly due to free-free emission. 


\subsection{Pathways for SN~1a explosions and novae}
One of the pathways for SN1a explosions, within the single-degenerate channel, is the 
double-detonation sub-Chandrasekhar mechanism wherein the WD accretes from a 
helium donor \citep[][]{2014ARA&A..52..107M}. A detonation in the 
accreted helium shell can cause a secondary detonation of the carbon/oxygen white dwarf 
(C/O WD) core \citep{2021gacv.workE..30S, 2010A&A...514A..53F, 2009ApJ...699.1365S}.
This happens at a total mass below the Chandrasekhar limit and depends on two  
factors – first the formation of a detonation in the helium shell and second whether a 
successful detonation of the helium shell can detonate the core. V445~Pup, by virtue of 
having had a shell detonation, is thus the only object which can be used as a testbed for 
the above theory.  

A secondary detonation can be triggered in two different ways: either directly when the helium 
detonation shock hits the core/shell interface (“edge-lit”), or with some delay, after the shock 
has converged near the center \citep{2010A&A...514A..53F}. \citet{2010A&A...514A..53F, 2007A&A...476.1133F}
examined the delayed mechanism since it could lead to a core detonation even for shocks 
too weak for the edge-lit case. They find that 
secondary core detonations are triggered for all their simulated models, ranging in core mass 
from 0.810\Msun{} up to 1.385\Msun{}  with corresponding shell masses from 0.126\Msun{} 
down to 0.0035\Msun. For convenience, we reproduce from Table~1 of  \citet{2010A&A...514A..53F}, 
the core mass of the WD and shell mass  in the format ($M_{\rm{core}}$, $M_{\rm{shell}}$) for all 
their models that undergo double detonation:  (0.81, 0.126), (0.92, 0.084), 
(1.025, 0.055), (1.125, 0.039), (1.28, 0.013), (1.385, $3.5 \times 10^{-3}$). The end result 
of their modeling is that  as soon as a detonation occurs in a helium shell covering a carbon/oxygen
WD a subsequent core detonation is virtually inevitable \citep{2010A&A...514A..53F}. 

The WD mass is unknown but we speculate it is low based on the low amplitude outburst, 
the extremely long time to decline, the formation of dust, the amount of mass ejected,  the 
low excitation spectrum at outburst \citep{2003A&A...409.1007A}, and the lack of coronal 
line emission even 3~yrs after outburst \citep{2004AJ....128.2962L}. 
\citet{2014MNRAS.445.3239P} suggest that the initial mass of the WD in
 V445~Pup could have been close to 0.8~\Msun. A low mass could 
explain why a SN~1a explosion was averted by the double-detonation channel.

The 2000 eruption of V445~Pup was not a SN explosion.
With $m_{V}^{\rm{max}} = 8.46$ \citep{2010PZ.....30....4G}, $A_{V} = 1.6$, $d$ in the range 3.5 to 8.2~kpc, 
the absolute magnitude $M_{V}$ is in the range $-6$ to $-7.7$, typical of a slow nova. Thus, 
its peak brightness falls greatly short of even the weakest SN “impostor” explosions which 
have $M_{V} \sim -13$ to $-14$ \citep{2012PASA...29..482K, 2009ApJ...697L..49S}. 

\subsection{V445 Pup -- Its nature and connection with SNe}
\begin{figure}[hb]
\figurenum{3}
\begin{center}
 \includegraphics[trim=1.5cm 8.2cm 5.5cm 8.0cm, clip, width=0.4\textwidth]{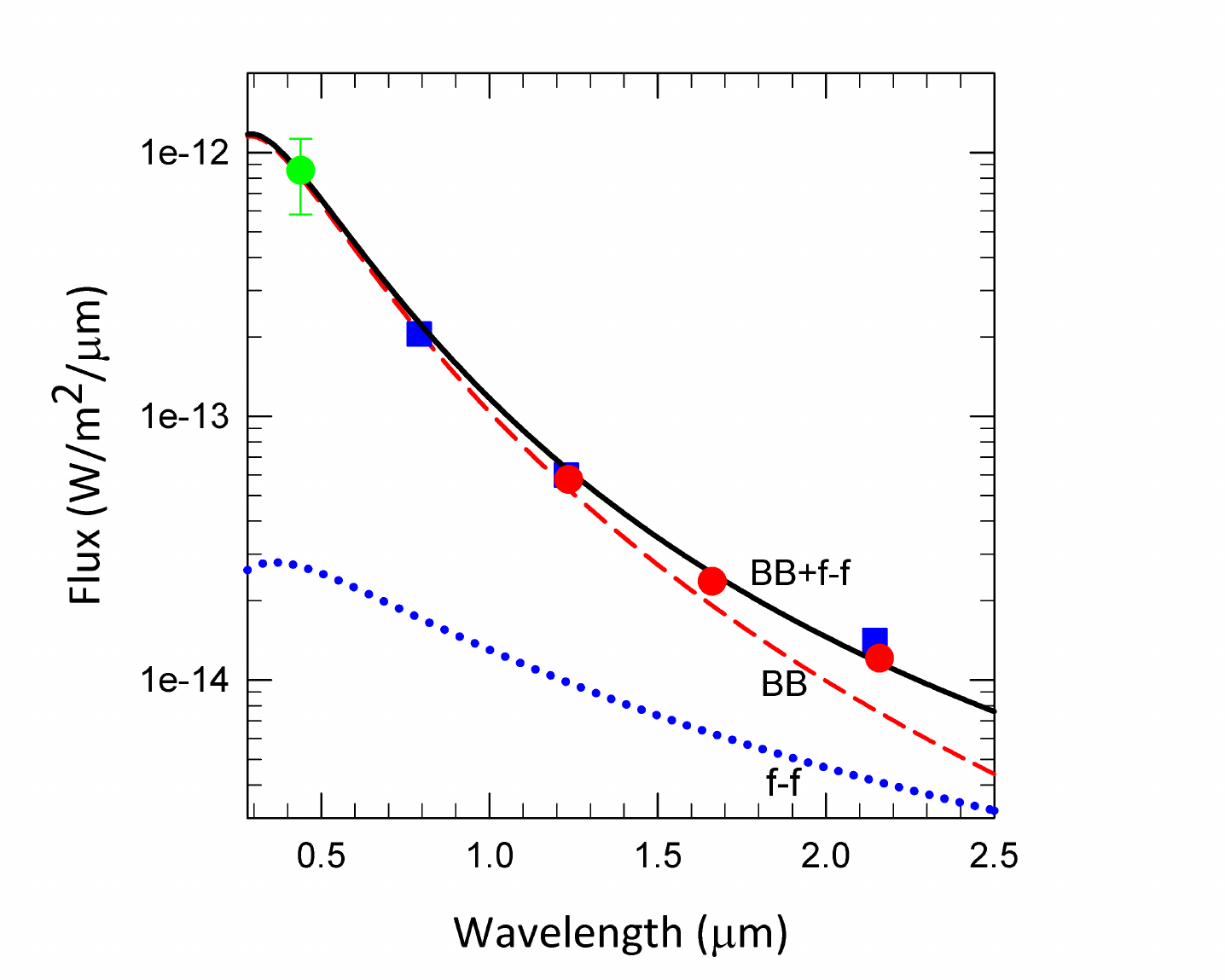}
\caption{The pre-eruption SED of the progenitor fit by 
the sum of a blackbody (BB) at 10000~K (dashed line) and free-free (f-f, dotted line) 
emission at 10000~K. The sum of the  BB and f-f emission is shown by the bold line.} 
\label{fig:sed-progen} 
\end{center}
\end{figure}

Useful insight about the system can be obtained from Fig.~\ref{fig:sed-progen} which 
presents a plot of the extinction corrected pre-outburst SED of the progenitor 
from 2MASS \citep{2006AJ....131.1163S} and DENIS \citep{1994ExA.....3...73E} magnitudes, 
both of which were recorded almost simultaneously 
on 1999 February 02 and 11 respectively  (2MASS: $J$ = 12.271 $\pm$ 0.026,  
$H$ = 11.94 $\pm$  0.024, $K_{s}$ = 11.52 $\pm$ 0.025; DENIS:  $I$ = 12.85 $\pm$ 0.02, 
$J$ = 12.24 $\pm$ 0.06, $K_{s}$ = 11.36 $\pm$ 0.09). We also use $B$ = 14.3 $\pm$ 0.3 from \citet{2010PZ.....30....4G}.
The SED is well fit by a $10^{4}$~K star, \citep[cf.,][]{2010PZ.....30....4G} but with a
discernible IR excess. This excess may be explained by free-free emission from ionized 
circumbinary gas that existed before the eruption (Fig.~\ref{fig:sed-progen}).


Several  factors point towards the existence  of material around V445~Pup prior to the
eruption. First, the radio synchrotron emission requires shocks formed by nova ejecta plowing 
into pre-existing material. Second, the bipolar morphology of the nebula necessitates an 
equatorial constriction for shaping \citep[the radio images confirm the constriction;][]{2021MNRAS.501.1394N}. 
Third, the putative presence of free-free emission demands ionized gas in the vicinity of 
the central star. Fourth, the presence of a large number of narrow absorption features 
(e.g.,  Ba~II 493.408~nm, Sc~II 552.679 nm and several lines of Ti ~II, Cr~II and Fe~II) seen 
in the rich, high dispersion ($R = 8000$) spectra \citep{2008A&A...482..865I} likely are
Transient Heavy Element Absorption (THEA) systems \citep{2008ApJ...685..451W}.
The latter authors propose that gas causing the THEA absorption systems in novae must 
be circumbinary, exists before the outburst, with a likely origin arising from mass ejection 
from the secondary star. 

Hence, preexisting (equatorial) material, if dusty, would contribute to the hotter (less massive) 
dust component. Our analysis does not rule out this possibility. Colder dust, at the temperature 
of the cooler component, may also have pre-existed  but in quantities below that required 
to allow a  pre-outburst detection by IRAS  \citep{1984ApJ...278L...1N} which had an 
average 10$\sigma$ sensitivity of 0.85~Jy at 60~\micron{} and 3~Jy at 100~\micron.

These  factors, when viewed collectively,  would favor  a pre-eruption configuration for V445~Pup 
that consists of a binary system, with remnants of the common envelope (CE) phase forming 
a torus  in the equatorial plane.  Radiation from a hot source, likely the WD or a hot accretion 
disc (or a combination of both) ionized parts of the CE remnant material leading to the 
observed free-free emission. The secondary was likely a star with $T_{eff} = 10^{4}$~K,
that was a periodic variable star with a probable orbital period of 
$0. 650654 \pm 0.000011$~d \citep{ 2010PZ.....30....4G}.

These parameters (spectral type, orbital period of secondary, mass of the 
ejecta) are essential inputs while modeling the thermal response of  a degenerate 
C/O WD accreting helium from a helium-star donor \citep[e.g.,][]{2016ApJ...821...28B, 2014MNRAS.445.3239P}. 
Both these cited studies show that various outcomes are possible from the accretion
process depending on the accretion rate and WD mass, whether there is steady He burning 
on the surface, mild shell flashes, strong shell flashes, or quiet accumulation of matter 
up to the final SN~1a explosion when the mass crosses the Chandrashekar limit.  If the 
accretion rate is low ($\dot{M} \ltsimeq 10^{-6}~\Msun~\rm{yr}^{-1}$) helium flashes 
result \citep[][]{2016ApJ...821...28B, 2014MNRAS.445.3239P} yielding a helium nova 
\citep{2019MNRAS.487.2538J}. 

Furthermore, helium novae could be related to Type~1ax SNe 
because the helium emission in some Type~1ax SNe appears to arise from the 
circumstellar environment rather than from the supernova ejecta itself \citep{2019MNRAS.487.2538J}. 
The argument is that the ejecta of a SN~1ax detonation,  following an earlier helium nova 
eruption on the same star, entrains the helium injected by the latter’s eruption into the 
circumstellar environment. Type~1ax supernovae share similar characteristics as SN~1a but 
exhibit  lower peak luminosities and ejecta velocities.  \citet{2019MNRAS.487.2538J} point 
out that the helium emission in two SN~1ax, SNe 2004cs and 2007J, is consistent with 
coming from the ejecta of a relatively recent helium nova, and note in particular, that the 
velocity of the material in these two SNe is similar to that of the galactic helium nova V445~Pup. 
Recently, \citet{2023Natur.617..477K}  discuss the strong likelihood of a V445~Pup type 
object being the progenitor of the first radio-detected Type1a SN~2020eyj which has  a helium-rich 
circumstellar medium. Explaining the radio light curve and the bolometric light-curve tail of SN2020eyj
requires a circumstellar medium mass between 0.3--1.0~\Msun. A mass of 0.67~\Msun, 
in good agreement with that posited for SN~2020eyj,  can be provided by the helium donor in 
V445~Pup if we adopt a distance of 8.2~kpc \citep{2009ApJ...706..738W} instead of 6.2~kpc,  
a plausible gas-to-mass ratio of 200, in tandem with the cold component mass of  $(1.9 \pm 0.8) \times 10^{-3}$ \Msun{} 
that we derive. V445~Pup is thus a unique test platform for testing single-degenerate channel 
theories that involve a helium-rich donor.  

Although the ejected mass in V445~Pup is unusually high for a nova, a helium nova 
outburst is still the most favorable interpretation for the 2000 eruption.  However, 
V445~Pup also shares certain similarities with CK~Vul, the latter proposed to belong to 
the class of objects known as intermediate-luminosity red transients (ILRTs) or 
interchangeably Luminous red novae \citep[][]{2020ApJ...904L..23B}.
CK~Vul also has a hourglass morphology, a  similar dust mass of $4.3 \times 10^{-3}$~\Msun{}
in the inner nebula, and peak expansion velocities of  $\simeq 2000~\rm{km~s}^{-1}$
 \citep{2020ApJ...904L..23B, 2018MNRAS.481.4931E}. However, CK~Vul was not 
 hydrogen deficient and was much more luminous at the peak of its outburst  
 ($M_{V} \sim -12.4$) compared to V445~Pup ($M_{V} \sim -7$).
 
 \section{Summary}
\label{sec-sum}
In case the bulk  of the estimated dust mass surrounding V445 Pup arose from material 
ejected in  the 2000 eruption, then  V445~Pup at the time of its outburst had a shell mass 
as massive as 0.01~\Msun{} or more and thus should have potentially undergone a 
SN~1a detonation by the double-detonation sub-Chandrasekhar pathway. 
That it did not suggests that  the WD is not massive. 
 
\begin{acknowledgments}
The authors thank the anonymous referee's critique of the manuscript that improved the
scientific narrative. This work is based in part on observations made with the Spitzer Space Telescope, 
obtained from the NASA/ IPAC Infrared Science Archive (doi: 10.26131/IRSA433), both of 
which are operated by the Jet Propulsion Laboratory, California Institute of Technology under a contract with 
the National Aeronautics and Space Administration. This publication makes use of data 
products from the Wide-field Infrared Survey Explorer (doi: 10.26131/IRSA142), which is a joint project of the 
University of California, Los Angeles, and the Jet Propulsion Laboratory/California Institute 
of Technology, funded by the National Aeronautics and Space Administration. Herschel is an 
ESA space observatory with science instruments provided by European-led Principal 
Investigator consortia and with important participation from NASA (doi: 10.26131/IRSA79). 
This research is based on observations with AKARI (doi:10.26131/IRSA180, 10.26131/IRSA181 ), 
a JAXA project with the participation of ESA. This publication makes use of data products from the 
Two Micron All Sky Survey (doi: 10.26131/IRSA2), which is a joint project of the University of Massachusetts and the 
Infrared Processing and Analysis Center/California Institute of Technology, funded by the 
National Aeronautics and Space Administration and the National Science Foundation and
the Deep Near-Infrared Survey of the Southern Sky (DENIS) Catalog (doi: 10.26131/IRSA478).

\end{acknowledgments}


\facilities{AKARI, 2MASS, DENIS, WISE, Spitzer, SEST, Herschel}

\software{IRAF, Astrophy \citep{2018AJ....156..123A}}

\clearpage

\end{document}